\documentclass[
    ,final            
  ]
  {aipproc}
\layoutstyle{6x9}
\usepackage{hyperref}

\begin{document}

\title{Composite Higgs models, Dark Matter and $\Lambda$}

\classification{12.60.-i,11.10.Kk,95.35.+d}
\keywords      {Electroweak symmetry breaking, Higgs models, dark matter}

\author{J. Lorenzo Diaz Cruz}{
  address={Facultad de Ciencias F\'isico-Matem\'aticas, Benem\'erita Universidad Aut\'onoma de Puebla, Puebla, Pue., CP: 72570, M\'exico}
}

\begin{abstract}
We suggest that dark matter can be identified with a stable composite fermion 
$X^0$, that arises within the holographic AdS/CFT models, where the
Higgs boson emerges as a composite pseudo-goldstone boson.
The predicted properties of $X^0$ satisfies the cosmological bounds,
with $m_{X^0} \sim 4\pi f\simeq O(TeV)$.  Thus, through a deeper 
understanding of the mechanism of electroweak symmetry breaking, 
a resolution of the Dark Matter enigma is found.
Furthermore, by proposing a discrete structure of the Higgs vacuum,
one can get a distinct approach to the cosmological constant problem. 
\end{abstract}

\maketitle


\section{INTRODUCTION}

The notion of spontaneous symmetry  breaking (SSB) \cite{Higgetal}
has been an important ingredient for the development of modern 
particle physics, with applications that range from the 
description of chiral symmetry breaking in the strong 
interactions \cite{GellLevi}, to the generation of 
masses in the electroweak model \cite{Glashow,WeinbSalam},
including as well the development of inflationary models \cite{Guth}.
However, the effective description of the Higgs mechanism 
still lacks a fundamental understanding.
 Thus, explaining the nature of electroweak symmetry breaking 
(EWSB) is one of the most important questions in particle physics 
today. Within the standard model (SM), electroweak precision
tests (EWPT)  prefer a Higgs boson mass of the order of 
the electroweak (EW) scale $v\simeq 175$ GeV \cite{HixOdissey}.  
Plenty of phenomenological studies have provided us with an understanding of
the expected properties of the Higgs boson (mass, decay rates and production
cross sections), which should be tested soon at the LHC. 

On the other hand, from the cosmology side, there are also big problems,
one of them being the enigma of dark matter (DM) \cite{COSMOrev}. 
 Plenty of astrophysical and cosmological data requires the existence 
of a DM component, that accounts for about 10-20\% 
of the matter-energy content of our universe \cite{DMrev}. 
A weakly-interacting massive particle (WIMP), with a mass also of the order
of the EW scale, seems a most viable option for the DM.
 What is the nature of DM and how does it fit into
our current understanding of elementary particles, is
however not  known.

Given the similar requirements on masses and interactions for both
particles, Higgs boson  and DM, one can naturally ask whether they 
could share a common origin. Within the minimal SUSY SM \cite{mssmrev}, 
which has become one of the most popular extensions of the SM, 
there are several WIMP candidates (neutralino, sneutrino, gravitino) 
\cite{mssmXDM}. Among them, the neutralino has been most widely studied; 
it is a combinations of SUSY partners of the Higgs and gauge bosons, the 
Higgsinos and gauginos. Thus, in SUSY models the fermion-boson symmetry 
provides a connection between the Higgs boson and DM.
However, many new models have been proposed more recently \cite{ewsbrev},
which provide alternative theoretical foundation to stabilize the Higgs 
mechanism. 
Some of these models, which have been originally motivated by 
the studies of extra dimensions \cite{exdimrev}, include new
DM candidates, such as the lightest T-odd particle (LTP) within 
little Higgs models \cite{LTP}  or the lightest KK particle (LKP) 
in models with universal extra-dimensions \cite{LKP}.

Here, we summarize the results of our search for possible dark
matter candidates, within the Holographic Higgs models \cite{myDMPRL}. 
In these constructions, EWSB is triggered by a light composite 
Higgs boson, which emerges as a pseudo-goldstone boson 
\cite{HoloHiggs,ADCHN}. 
Within this class of models, we propose that a stable  composite 
``Baryon'', tightly bounded by the new strong interactions, 
can account for the DM.
 This picture, where strong interactions produce a light pseudo-goldstone
boson and a heavier stable fermion, is not strange at all in nature.
 This is precisely what happens in ordinary hadron physics, where
the pion and the proton play such roles. In this paper we shall discuss 
models that produce a similar pattern for the Higgs and DM, but at a higher 
energy scale, and with a stable neutral state instead of a charged one. 

However, even if the Higgs bosons is found at the LHC, and even if one
could identify the Dark matter candidate,
there will be some issues left open.  One of them, probably the most 
difficult one, is the cosmological constant problem. Namely, we would like
to understand why the Higgs vacuum does not produce the large curvature
that one would expectw with naive estimates.  Many efforts have been
devoted to this problem, but so far no solution has been found.
This issue would probably need an understanding of the structure of 
space-time \cite{SpacetimeStr}.

Here, we also present our discrete model of the Higgs vacuum 
\cite{OurDiscrH}, which departs from the usual continuum model. 
Namely we shall assume that the Higgs vacuum has an structure, 
and it consists of small size regions (droplets) 
where the vacuum expectation value is different from zero,
while in the true empty regions it vanishes. For simplicity we shall
consider that these regions form spherical droplets, and it will be 
shown that this model allows to solve the cosmological constant
problem,  for a certain relation between the density and size of the
sherical droplets. The model is not distinguishable from the SM at
the energies of current accelerators, however interesting deviations
can be expected to  occur at the coming LHC or higher energies.

\section{Holographic Higgs Models and Dark Matter}

The Holographic Higgs 
models of our interest, admit a dual AdS/CFT description, however, we shall 
discuss its features mainly from the 4D point of view, using first a 
generic effective lagrangian approach,  and then presenting specific 
realizations within the known  Holographic Higgs models \cite{HoloHiggs,ADCHN}.
 From the 4D perspective, the effective lagrangian that describes 
these models \cite{EffholoHIX,EffholoKK}, includes two sectors: 
i) The SM sector that contains the gauge bosons and 
most of the quarks and leptons, which is characterized by a 
generic coupling $g_{sm}$ (gauge or Yukawa), 
and ii) A new strongly interacting sector, characterized
by another  coupling  $g_{*}$ and an scale $M_R$. This scale can be associated
with the mass of the lowest composite resonance, which in the dual
AdS/CFT picture corresponds to the lightest KK mode; in ordinary QCD 
$M_R$ can be taken as the mass of the rho meson ($\rho$). 
 The couplings are choosen here to satisfy  $g_{sm} \sim g_{*} \sim 4 \pi$, 
and as a result of the dynamics of the strongly interacting sector, 
a composite Higgs boson emerges. It behaves as an  exactly 
massless goldstone boson because of the global symmetries that hold
in the limit $g_{sm} \to 0$. 
SM interactions then produce a deformation of the theory, and the
Higgs boson becomes a psudo-Goldstone boson. Radiative effects induce a 
Higgs mass, which can be written as:
$m_h \simeq (\frac{g_{sm}}{4\pi}) M_R$.

Simultaneous to the Higgs appearance, a whole tower of fermionic 
composite states $X^0, X^{\pm}, X^{ \pm\pm}...$ should also appear.
Our dark matter candidate is identified with the lightest
neutral state ($X^0$) within this fermionic tower, and we call it
the lightest Holographic fermionic particle (LHP for short).
Similarly to what happens in ordinary QCD, where the proton is 
stable because of Baryon number conservation, we also assume 
that $X^0$ is stable because a new conserved quantum number, 
that we call  ``Dark Number'' ($D_N$). 
Thus, the SM particles and the ``Mesonic'' states, like 
the Higgs boson,  will have zero Dark number ($D_N(SM)=0$), 
while the ``baryonic'' states like $X^0$,  will have +1
dark number ($D_N(X^0)=+1$). The formation of such ``baryonic'' states,
including a conserved number of topological origin, has been derived 
recently using the Skyrmion model in the RS geometry \cite{dualSkyrme}.
 For a  strongly interacting sector that corresponds to a deformed  
$\sigma$ type model, the mass of $X^0$ satisfies:  $M_{X^0} \sim 4 \pi f$, 
where $f$ is the analogue of the pion decay constant, 
thus $m_{X^0} \simeq M_R$.  
In analogy with ordinary QCD, it is usualy assumed that lightest
resonance  corresponds to a vector meson, however $X^0$ itself could be
the lightest state. In any case, the natural value for $M_{X^0}$ 
will be in the TeV range,  somehow heavier than the SUSY candidates 
for DM. It is important to stress that because  $\Lambda_H \simeq M_R$, then
the EWPT analysis can be reinterpreted as an indirect method to 
obtain constraints on the dark matter scale.

There are several alternatives 
to accomodate our proposed LHP candidate, within the Holographic 
Higgs models proposed so far \cite{HoloHiggs}, and it is one of the
purposes of our work to identify the most favorable models.
From the 4D perspective, each model is defined by impossing
a global symmetry $G$ on the new strongly interacting sector, 
then a subgroup $H$ of $G$ will be gauged; here we shall consider the
case when the SM group is gauged, i.e. $H=SU(2)_L \times U(1)_Y$.
Furthermore, in order to fix the LHP  quantum numbers, one needs to
specify a particular representation ($G$-multiplet) that will contain it.  
Then, this $G$-multiplet can be decomposed in terms of an $H$-multiplet 
plus some extra states. We call {\it{Active DM}} those cases 
when the LHP belongs to the $H$-multiplet,  while {\it{Sterile  DM}} 
will be used for models where the LHP is a SM singlet. 

Let us consider first the models based on the group
$G=SU(3) \times U(1)_X$ \cite{HoloHiggs}. $U(1)_X$ is needed
in order to get the correct SM hypercharges. Under 
$SU(3) \times U(1)_X$ the SM doublets ($Q$) and d-type singlets ($D$)  
are included in  $SU(3)$ triplets, i.e.  
$Q \equiv \mathbf{3}^*_{1/3}$, $D \equiv \mathbf{3}_0$. 
The SM up-type singlet ($U$) is defined as a TeV-brane singlet field,
i.e. $U \equiv \mathbf{1}_{1/3}$. The hypercharge is
obtained from: $Y=\frac{T_8}{\sqrt{3}}+X$, while
the electric charge arises from:  $Q_{em}=T_3+ Y$, 
and $T_{3,8}$ denote the diagonal generators of $SU(3)$.
Then, admiting only the lowest dimensional  $SU(3)$ representations 
(triplets and singlets), one can obtain the electrically 
neutral LHP, by requiring: $X= \pm 1/3, \pm 2/3$. 
Thus, for an $SU(3)$ anti-triplet with $X=1/3$: 
$\Psi_1= (N^0_1, C^+_1,N^0_2)^T$, there are two options for the LHP: 
i) Model 1 (active): the LHP belongs to a SM doublet $\psi_1=  (N^0_1, C^+_1)$, 
i.e. $X^0=N^0_1$, and 
ii) Model 2 (sterile): the LHP is a SM singlet, i.e. $X^0=N^0_2$.
Similar pattern is obtained for $X=-1/3$.
Choosing instead a $SU(3)$ triplet with $X= \pm 2/3$, i.e.
$\Psi_2= (N^0_3, C^+_2,C^+_3)^T$, only allows the LHP to be 
$X^0=N^0_3$ (Model 3).
 Allowing the inclusions of SU(3) octets leads to the possibility of
having LHP candidates that belong to SM triplets with $Y=0, \pm 1$ 
(Models 4,5). 

On the other hand, LHP candidates can also arise
within the minimal composite Higgs model ($MCHM$) with global
symmetry  $G=SO(5) \times U(1)_X$ \cite{holohixM}, 
which incorporates a custodial symmetry. 
The SM hypercharge is defined now by 
$Y=X+T_3^R$, where $T^R_3$ denotes the R-isospin obtained from the 
breaking chain:
$SO(5)\times U(1)_X \to SO(4)\times U(1)_X \to SU(2)_L\times U(1)_Y$,
and with $SO(4) \simeq SU(2)_L\times SU(2)_R$.
In the model $MCHM_5$, the SM quarks and leptons are accomodated in the 
fundamental representations ($5$) of $SO(5)$, while in the option named  
$MCHM_{10}$, the SM matter is grouped in the anti-symmetric ($10$-dimensional) 
representation of $SO(5)$. For the DM candidates one can use 
either of these possibilities. 
DM models using the $5$  of $SO(5)$, 
can accomodate the LHP in SM doublets or singlets, similar to
the pattern obtained for the $SU(3)$ models. 
On the other hand, in models that employ the $10$
representation of $SO(5)$, the LHP can also appear in SM triplets.
 For instance, taking $X=0$,  allows $X^0$ to fit in a $Y=0$ triplet, 
while the option $X=\pm 1$, offers the posibility of having an LHP
within  a $Y=\pm 1$ triplet.

The effective lagrangian description of both the Higgs and DM,
is given by:
\begin{equation}
{\cal{L}}_H = {\cal{L}}^H_{sm}+{\cal{L}}_{DM} + \sum \frac{\alpha_i}{(\Lambda_H)^{n-4}}O_{in},
\end{equation} 
where ${\cal{L}}^H_{sm}$ denotes the SM Higgs lagrangian.
 The higher-dimensional operators $O_{in}$ ($n\geq 6$) can induce 
corrections to the SM Higgs properties; meassuring these effects at 
future colliders (LHC,ILC), could provide information on the DM scale. 
The coefficient $\alpha_i$ and the scale $\Lambda_H$ will depend 
on the nature of each operator. The leading operators are: 
$O_W=i(H^\dagger \sigma^i D^\mu H)(D^\nu W_{\mu\nu})^i$,
$O_B=i(H^\dagger D^\mu H)(\partial^\nu B_{\mu\nu})$,
$O_{HW}=i(D^\mu H)^\dagger \sigma^i (D^\nu H) W^i_{\mu\nu}$,
$O_{HB}=i(D^\mu H)^\dagger (D^\nu H) B_{\mu\nu}$,
$O_{T}=i(H^\dagger D^\mu H)(H^\dagger D_\mu H)$,
$O_H =i \partial^\mu (H^\dagger H) \partial_\mu (H^\dagger H)$
\cite{EffholoHIX}.
At LHC it will be possible to meassure the corrections 
to the Higgs couplings, with a precision that will 
translate into a bound $\Lambda_H \geq 5-7$ TeV, while
at ILC it will extend up to about 30 TeV \cite{EffholoHIX}. 
These operators can also modify the  SM bounds 
on the Higgs mass obtained from EWPT. In particular,
$O_T$ can increasse the limit on the Higgs mass 
above 300 GeV, for $\alpha_i=O(1)$ and $\Lambda_H \simeq 1$ TeV. 

 The renormalizable interactions of $X^0$ with the SM, are fixed by
its quantum numbers, while the complete effective lagrangian includes
 higher-dimensional operators, namely:
\begin{equation}
{\cal{L}}_{DM} = \bar{X}^0 (\gamma^\mu D_\mu - M_X) X^0
+ \sum \frac{\alpha_i}{(\Lambda_X)^{n-4}} O_{in}
\end{equation} 
where $D_\mu= \partial_\mu -i g_{x} T^i W^i_\mu- g'_{x} \frac{Y}{2} B_\mu$.
For those operators that describe composite effects, one expects that 
$\Lambda_X \simeq f$, while for operators that result from the
integration of the G-partners of $X^0$, one expects 
$\Lambda_X \simeq M_R > M_X$. Similarly, the coupling $\alpha_X$
should be of order $O(1)$ ($b_i/16 \pi^2$) for operators induced
at tree- (loop-) level.

We are interested in constraining the LHP models,
using both cosmology (relic density) and the experimental searches for DM.
We shall consider the three types of models:
i) Active LHP models with $Y\neq 0 $, ii) Active LHP models with $Y=0$,
and iii) Sterile LHP models. Let us discuss first the active LHP models. 
The corresponding relic density  can be written in terms 
of the thermal averaged cross-section $<\sigma v>$ as follows: 
\begin{equation} 
\Omega_X h^2 = \frac{2.57\times 10^{-10} }{ <\sigma v>}=
\frac{ 2.57\times 10^{-10} M^2_X}{C_{T,Y} }  
\end{equation} 
where $C_{T,Y}$ depends on the isospin (T) and
hypercharge (Y) of the LHP. Numerical values of
$C_{T,Y}$ for the lowest-dimensional representations are:  
$C_{1/2,1/2}=0.004 $,  $C_{1,0}= 0.01$,  $C_{1,1}= 0.011$.
Then, in order to have agreement with current data, 
i.e. $\Omega_X h^2 =0.11\pm 0.066$ \cite{wmapA},
models 1,3  require $M_X=1.3$ TeV, while model 4 (5)
require $M_X=2.1$ ($M_X=2.2)$ TeV, respectively. It is quite remarkable 
that these values are precisely of the right order expected in the 
strongly interacting Higgs model!.

In order to discuss the relic density constraint for the sterile 
LHP DM (model 2), we notice that the couplings of $X^0$ with the SM 
gauge and Higgs bosons, come from the higher-dimensional operators, 
which include i) 4-fermion operators:
$O^1_{FX} = \frac{1}{2} (\bar{F} \gamma^\mu F) ( \bar{X} \gamma_\mu X)$,
$O^1_{fX} = \frac{1}{2} (\bar{f} \gamma^\mu f) ( \bar{X} \gamma_\mu X)$,
$O^V_{FX} = \frac{1}{2} (\bar{F} \gamma^\mu X) ( \bar{X} \gamma_\mu F)$,
$O^V_{fX} = \frac{1}{2} (\bar{f} \gamma^\mu X) ( \bar{X} \gamma_\mu f)$,
$O^S_{FX} = \frac{1}{2} (\bar{F}  X) ( \bar{X}  F)$,
$O^S_{fX} = \frac{1}{2} (\bar{f}  X) ( \bar{X}  f)$,
ii) fermion-scalar operator: $O_{X\phi}= (\Phi^\dagger \Phi)  ( \bar{X}  X)$,
and iii) Fermion-vector-scalar operator:
$O_{DX}= (\Phi^\dagger D^\mu \Phi)  ( \bar{X} \gamma_\mu X)$.
where $F(f)$ denote the SM fermion doublet (singlet). The full analysis
should include all these operators, which depends on many
parameters, however, to obtain a simplified estimate,
we shall only consider the operator $O_{DX}$. 
This operator induces an effective vertex $Z X^0 X^0$ of
the form: $\Gamma_{ZXX}= \frac{g}{2c_W} \eta \gamma^\mu$, 
with $\eta=2 c_x g c_w v^2/M^2_R$, and $c_x$ being the 
coefficient of $O_{DX}$. Then, requiring 
$\Omega_X h^2 \simeq \Omega_{DM} h^2 = 0.11\pm 0.006$ \cite{wmapA}, 
implies: $M_X \simeq 0.8 \eta$ TeV. Thus, for
$M_X$ of order TeV, one would need to have $\eta \geq 1$, which
could be satisfied in some region of parameter space, although
one usually expects $\eta \leq 1$ within a strongly interacting 
scenario. 

Constraints on the LHP models can also be derived from the 
direct experimental search for DM, such as the one based on the 
nucleon-LHP elastic scattering \cite{CDMexp}.  The corresponding 
cross section can be expressed as: 
$ \sigma_{T,Y} = \frac{G^2_F}{2\pi} f_N Y^2$, 
where $f_N$ depends on the type of nucleus used 
in the reaction. As it was discussed in ref. \cite{MinDM}, 
vector-like dark matter with $Y=1$ is severely constrained by the
direct searches, unless its coupling with the Z boson 
is suppressed with respect to the SM strength.
A suppression of this type can be realized in a natural manner for 
Holographic DM  models. Namely, following ref. \cite{EffholoKK}, 
we notice that by admitting a mixing between the composite LHP
and a set of elementary fields with the same quantum numbers,
then the vertex $ZXX$ will be suppressed by the mixing angles
needed to go from the weak- to the mass-eigenstate basis. 
For model 1, with active DM appearing in a doublet 
$\psi_1=  (N^0_1, C^+_1)^T$, one includes 
an elementary copy of these fields, which then
allows to write the vertex $ZXX$ as:
$\Gamma_{ZXX}= \frac{\eta' g_2}{2c_W} \gamma^\mu$, 
with $\eta' <1$. The cross-section for
$DM+N \to DM+N$ can be written then as:
$\sigma = \frac{G^2_F}{2\pi} f_N {\eta'}^2$.
Agreement with current bounds \cite{CDMexp} requires
to have ${\eta}'^2 \leq 10^{-2}-10^{-4}$, which seems reasonable.
On the other hand, DM with $Y=0$ automatically satisfies this bound, i.e.
$\sigma(Y=0)=0$. While for sterile dark matter, the
corresponding nucleon-LHP cross-section, 
satisfies the current limits \cite{CDMexp}, provided that 
the factor $\eta$ also satisfies ${\eta}^2 \leq 10^{-2}-10^{-4}$,
which is in contradiction with the bound derived from the cosmological
relic density, i.e. $\eta \geq 1$, therefore we find that the sterile
dark matter candidate (Model 2) seems disfavored.

\section{ A Discrete model of the Higgs vacuum}
Our presentation of
the Higgs mechanism starts by considering  an scalar field
that interacts with gauge bosons and fermions. The lagrangian for
the scalar and gauge sectors is written as:
\begin{equation}
{\cal{L}}= (D^{\mu} \Phi)^{\dagger} D_{\mu} \Phi - V(\Phi)
\end{equation}
where the covariant derivative is given by:
$D_{\mu} \Phi = (\partial_\mu -ig T^a V^a_\mu) \Phi$,
and $T^a$ are the generators for the representation that $\Phi$
belongs to. The Higgs potential takes the ``Sombrero Mexicano''
form, which has a minimum at a value of the Higgs field
$<\Phi>= v$, which is assumed to accur everywhere.

As we discussed in our paper \cite{OurDiscrH}, the continuum Higgs 
v.e.v. will be replaced by a distribution, i.e. we shall assume that the 
Higgs v.e.v. is
different from zero only in some small regions (droplets), elsewhere
the v.e.v. will be zero. Thus, we take the view 
that the Higgs vacuum is really a Bose condensate. 
Such condensates have been studied in condensed matter,
where certain compounds are made of certain atoms, e.g. Helium, 
that favor the emergence of such phenomena. Therefore, one could be 
tempted to extrapolate that such ``molecular'' or``atomic'' structures 
should also exist in order to explain the true nature of the Higgs
mechanism. In this paper, we shall consider that this may be a
possibility, but will leave open the possibility that our ``droplets'' 
are indeed those ``atoms'' or a ``molecule'' or a larger collection of 
such atoms, which will define a hierarchy of scales.   

If the vacuum energy (v.e.v.) were spread continuosly, it
would contribute to the cosmological constant, with a value of the
order $\Lambda \simeq 10^9$ GeV$^4$ = $10^{49}$ GeV/$cm^3$. However if
the droplets, of finite size $r_d$ and inter-distance $l_d$, 
are distributed uniformly with  a density $\rho_d$, 
then their contribution to the cosmological constant would
be $\Lambda = \rho_d v \tau_d$, where the volume of the droplets
is given by $\tau_d \simeq r^3_h$. 
Thus, by saturating the observed value for the cosmological constant 
($\simeq 10^{-4} GeV/cm^3$), we obtain  $\rho_d \tau_d \simeq 10^{-56}$. 
Furthermore, by considering that the distance between the droplets 
should be smaller than the shortest distance being tested at current 
colliders, i.e. $l_d \leq 10^{-15}$ cm, then the resulting
size of the droplest is of the order of the Planck length,
i.e. $r_d \simeq 10^{-33}$ cm. 

One may wonder why current experiments have not detected the
structure of the Higgs vacuum. The reason is that current probes
(photons, electrons, protons) have an energy or momentum that
corresponds to a wave-length that is large than the distance between
the spheres with v.e.v. different from zero. Thus, with current probes
the vacuum ``looks'' continuum. However, one one gets an energy that
is if the order of the inverse of the distance between the spheres,
the vacuum will start to show its structure.

In order to identify possible test of our model that can be carried  at
the LHC, we shall focus on Higgs phenomenology. 
Let us consider the standard Higgs interaction with a fermion $\psi$, which
is described by the Yukawa lagrangian. After SSB we get the mass
of the fermion and its interaction with the Higgs.
However, this will be valid only at low energies, but at high-energies 
the fermion will ``see'' less v.e.v., therefore the coupling will be
not be given just by the fermion mass, but rather we need to include an 
energy-dependent factor for the vertex and the mass:

\begin{equation}
{\cal{L}}_{new}  = x(q^2)\frac{m}{v} h  {\bar{\psi}}_L \psi_R +  
                    m(q^2)  {\bar{\psi}}_L \psi_R + h.c. 
\end{equation}

These effect can be probed at the LHC by looking at the Higgs
production. For instance we can study the gluon fusion production,
which depends on the Higgs coupling with the top. Now the
cross-section needs to include the form factor $x(q^2)$, which will
affect the shape of the $p_T$ distributions. At lower
momenta the result will be similar to the SM, but at higher momenta, 
we will observe a deviation from the SM result.

\section{Conclusions}

We have proposed a new DM candidates (LHP),  within the context of 
strongly interacting Holographic Higgs models. LHP candidates
are identified as composite fermionic states ($X^0$),  with a 
mass of order $m_{X^0} \sim 4\pi f$,  which is made stable 
by assuming the existence of a conserved ``dark'' quantum number. 
Thus, we suggest that there exists a connection 
between two of the most important problems in particles physics and 
cosmology: EWSB and DM.
 In these models, the Higgs couplings receive potentially 
large corrections, which could be tested at the coming (LHC) and 
future colliders (ILC). Measuring these deviations, could also provide 
information on the dark matter scale.  
 We have verified that the LHP relic abundance is satisfied for masses of 
$O(TeV)$, which is the range expected in Holographic Higgs models.  
Furthermore, the current bounds on experimental searches for DM 
based on LHP-nucleon scattering, provides further 
constraints on the possible models.  
Overall, we conclude that most favorable models are the active ones with
$Y=0$.  It could be interesting to compare our model with other
approaches tha predict a composite dark matter candidate
\cite{compoDM}.

Here, we have also presented a model of the Higgs vacuum, which
assumes that the Higgs vacuum has an structure, it consists of small size
regions where the vacuum expectation value is different from zero,
while in the true empty regions it vanishes. For simplicity we shall
consider that these regions form spherical droplets, and it will be 
shown that this model allows to solve the cosmological constant
problem,  for a certain relation between the density and size of the
sherical droplets. The model is not distinguishable from the SM at
the energies of current accelerators, however interesting deviations
can be expected to  occur at the coming LHC or at higher energies.

\begin{theacknowledgments}
The author thanks the {\it Sistema Nacional de Investigadores} and
{\it CONACyT} (M\'{e}xico) for financial support. The seond part of
this article is based on a collaboration with P. Amore and A. Aranda,
which I sincerely acknowledge.
\end{theacknowledgments}

\end{document}